\documentclass[%
 aip,
 amsmath,amssymb,
 reprint,%
]{revtex4-1}

\usepackage{graphicx}
\usepackage{dcolumn}
\usepackage{bm}

\usepackage[utf8]{inputenc}
\usepackage[T1]{fontenc}
\usepackage{mathptmx}
\usepackage{mathrsfs}
\usepackage{etoolbox}

\makeatletter
\def\@email#1#2{%
 \endgroup
 \patchcmd{\titleblock@produce}
  {\frontmatter@RRAPformat}
  {\frontmatter@RRAPformat{\produce@RRAP{*#1\href{mailto:#2}{#2}}}\frontmatter@RRAPformat}
  {}{}
}%
\makeatother

\usepackage{amsmath}
\usepackage{amsfonts,amssymb}
\usepackage{mathrsfs}
\usepackage{graphicx,graphics}
\usepackage{hyperref}

\usepackage{verbatim}
\usepackage[normalem]{ulem}

\usepackage{color}

\newcommand{\rtort}{r_{*}}
\newcommand{\ricci}{\mathcal{R}}
\newcommand{\flux}{\mathcal{F}_{\infty}}
\newcommand{\Laplace}{\mathscr{L}}
\newcommand{\meff}{m_{\xi}^{2}}

\begin{document}

\preprint{AIP/123-QED}

\title[]{Boundary conditions for isolated asymptotically anti-de Sitter spacetimes}
\author{Christyan C. de Oliveira}
 \email{chris@ifi.unicamp.br}
 \affiliation{Instituto de Física ``Gleb Wataghin" (IFGW), Universidade Estadual de Campinas, 13083-859, Campinas, SP, Brazil
 }
\author{Ricardo A. Mosna}%
 \email{mosna@unicamp.br}
\affiliation{ 
Departamento de Matem\'atica Aplicada, Universidade Estadual de Campinas, 13083-859, Campinas, SP, Brazil}%
\author{Jo\~ao Paulo M. Pitelli}
\email{pitelli@unicamp.br}
\affiliation{%
Departamento de Matem\'atica Aplicada, Universidade Estadual de Campinas, 13083-859, Campinas, SP, Brazil
}%

\date{\today}
             
\begin{abstract}
We revisit the propagation of classical scalar fields in a spacetime which is asymptotically anti-de Sitter. The lack of global hyperbolicity of the underlying background gives rise to an ambiguity in the dynamical evolution of solutions of the wave equation, requiring the prescription of extra boundary conditions at the conformal infinity to be fixed. We show that the only boundary conditions tha are compatible with the hypothesis that the system is isolated, as defined by the (improved) energy-momentum tensor, are of Dirichlet and Neumann types. 
\end{abstract}

\maketitle

\section{Introduction}
\label{sec:intro}
The anti-de Sitter (AdS)  spacetime is the classic solution to the vacuum Einstein equations in the presence of a negative cosmological constant. It has the highest possible degree of symmetry since it is maximally symmetric.  Despite this apparent geometric simplicity, the AdS spacetime has remarkable properties that make it a particularly interesting background for the study of classical and quantum fields.  In particular, it is a non-globally hyperbolic spacetime, implying that the solutions of the wave equation are not fully determined from initial data~\cite{wald1984general}. This requires the prescription of extra boundary conditions at its spatial infinity in order to have a unique solution for the Cauchy problem~\cite{wald1980dynamics}.  Physically, the lack of global hyperbolicity is related to the fact that information propagating in AdS can reach spatial infinity in finite time, which allows the energy to leak out of the spacetime. As a result, the AdS spacetime does not give rise, in general, to an isolated system.

This problem has been addressed in Refs. \onlinecite{breitenlohner1982positive,breitenlohner1982stability}  within the context of {\em supergravity} in (1 + 3)-dimensions.
Besides analyzing the stability of the anti-de Sitter background with respect to small scalar perturbations, these works show that the boundary conditions that make the improved energy functional positive and conserved are restricted to the Dirichlet and Neumann types.

Given the arbitrariness on the choice of the boundary condition at the conformal boundary,  Wald and Ishibashi defined in Refs.~\onlinecite{ishibashi2003dynamics, ishibashi2004dynamics,wald1980dynamics} a sensible prescription for obtaining the dynamics of a propagating field on AdS.%
\footnote{In fact, Refs.~\onlinecite{ishibashi2003dynamics, wald1980dynamics} deal with a general non-globally hyperbolic spacetime,  where the lack of global hyperbolicity arises due to a conformal boundary or a naked singularity.  Ref.~\onlinecite{ishibashi2004dynamics} works, specifically,  with the AdS spacetime.}
By requiring that the field propagation respects causality and time translation/reflection invariance and, what is most important, also has a conserved energy functional, it was shown that the non-equivalent types of sensible dynamics are in one-to-one correspondence with the positive self-adjoint extensions of the spatial part of the wave operator. These self-adjoint extensions are obtained by choosing suitable boundary conditions at the conformal infinity. The resulting conserved energy functional, however, is not that extracted from the improved energy-momentum tensor $T_{\mu\nu}$.  In fact, it can be shown  that it arises from the subtraction of a boundary term from the  energy functional  coming from $T_{\mu \nu}$.~\cite{avis1978quantum} This boundary term vanishes for Dirichlet or Neumann boundary conditions, and in this case, the newly defined (conserved) energy matches the usual energy, which is already conserved. For every other --- generalized Robin ---  boundary condition, there is an effective contribution of the boundary term to the newly defined (conserved) energy functional, showing that there is an effective flux of energy through the conformal boundary of AdS.

In any event, Robin boundary conditions have recently spawned great interest in the context of Quantum Field Theory in asymptotically anti-de Sitter spacetimes and several authors analyzed the consequences implied by these boundary conditions on the quantization of the scalar field (see, for instance, Refs.  \onlinecite{dappiaggi2016hadamard,dappiaggi2018superradiance,ferreira2017stationary,morley2020quantum,barroso2020boundary} and references therein). As a matter of fact, the introduction of Robin boundary conditions is often motivated by the desire that the system be isolated, as explicitly stated in Refs. \onlinecite{dappiaggi2018superradiance,ferreira2017stationary,barroso2020boundary}.
One of the goals of the present work is to clarify this issue and show that generic Robin boundary conditions are incompatible with the requirement that the spacetime be isolated.

More precisely, this paper is concerned with the Cauchy problem associated with the wave equation 
\begin{align}
(\Box - \meff)\Phi = 0
\label{eq:wave equation general}
\end{align}
in an \emph{asymptotically} anti-de Sitter spacetime, where $ \meff \equiv \mu ^ {2} + \xi \ricci $ and  $ \xi $ is a constant which couples the field  to the curvature scalar $ \ricci $. This coupling modifies the usual energy-momentum tensor obtained by the variation of the action with respect to the metric. In what follows, we use the resulting improved energy-momentum tensor to define the energy functional. Our aim is to establish the boundary conditions for which the system {\em spacetime} + {\em field} can be considered as effectively isolated, a point which, as mentioned above, has occasionally been a source of confusion in the literature.
It turns out that this is equivalent to finding the boundary conditions for which the conserved energy functional defined by Wald and Ishibashi is equal to the one extracted from the improved energy-momentum tensor. We emphasize that our analysis takes into account only classical fields.  In the context of quantum fields in curved spacetimes, the prescription of Wald and Ishibashi leads to a vanishing (renormalized) energy flux $\langle T_{t\rho}\rangle$ (see Ref.~\onlinecite{barroso2020boundary}).

This paper is organized as follows. In Sec. \ref{sec:Asymptotic behaviour of the field}, we obtain an asymptotic expression for the scalar field at spatial infinity. This is done by means of a Green function that encodes the dependence of the solution on the initial data and boundary conditions. Our analysis differs from that in Refs.  \onlinecite{breitenlohner1982positive,breitenlohner1982stability,ishibashi2004dynamics} in that we only assume that the spacetime is \emph{asymptotically} AdS; we thus make no assumption (except for certain technicalities to be explained below) about its bulk structure. In Sec. \ref{sec:Energy flux in asymptotically anti-de Sitter spacetimes}, we discuss the requirements on the boundary conditions at spatial infinity for the system {\em spacetime} + {\em scalar field} to be effectively isolated. We find that the only boundary conditions that are compatible with this assumption are the (generalized) Dirichlet and Neumann  boundary conditions. Finally, in Sec. \ref{sec:Discussion}, we discuss our results and make our closing remarks.

\section{Asymptotic behavior of the field}
\label{sec:Asymptotic behaviour of the field}

Let $M$ be a stationary $n$-dimensional spacetime, which is asymptotically AdS. We choose coordinates $ \{t,r, \theta_{1}, \dots, \theta_{n-2} \}$ such that the metric on $M$ satisfies
\begin{align}
ds^{2}|_{r\to \infty} \approx ds^{2}_{\!AdS} = -(1+r^{2})dt^{2} +  \nonumber\\
\frac{dr^{2}}{1+r^{2}} +r^{2} \, d\Omega^{2}_{n-2},
\label{eq:AdSn metric}
\end{align}	
where $ds^{2}_{AdS}$ is the line element in AdS$_{n}$ and $d\Omega^{2}_{n-2}$ is the metric on the $(n-2)$-dimensional unity sphere.

We separate variables for the scalar field and consider the {\em ansatz} 
\begin{align}
\Phi(t,r,\theta) = \sum_{\{\ell\}} \phi_{\ell}(r,t) Y_{\ell}(\theta),
\end{align}
where $\{\ell\}$ represents the set of integer indices labeling the {\em hyperspherical harmonics} $Y_{\ell}(\theta)$. The wave equation~\eqref{eq:wave equation general} can then be written as 
\begin{align}
L_{rt}[\phi]=0,
\label{eq:radial temporal equation}
\end{align}
where $L_{rt}$ is a second order differential operator of the form
\begin{align}
L_{rt} = u^{ij}(r)\partial_{i}\partial_{j} + v^{i}(r)\partial_{i} +q(r), \quad \quad & i,j=r,t.
\end{align}  

When dealing with problems such as \eqref{eq:radial temporal equation}, it is common practice to consider a time dependence of the form $ e ^ {- i \omega t} $ and then to solve the resulting time-independent problem. However, when considering non-conservative systems (for instance, when energy can flow through the boundaries), with $ \omega$ being a complex number, such an approach leads to extra mathematical difficulties, which, in turn, make it difficult to physically interpret the resulting solutions. \cite{nollert1992quasinormal,andersson1995excitation,berti2006quasinormal} When the spacetime bulk contains a black hole, such an approach allows for the determination of the quasinormal mode spectrum of the system. However, the quasinormal modes do not provide a complete set of eigenfunctions, and, hence, an arbitrary initial condition cannot be expressed in terms of them. 

As discussed in Ref. \onlinecite{nollert1992quasinormal}, one can overcome this difficulty by taking the initial conditions into account from the beginning. A suitable mathematical tool for implementing this strategy is the Laplace transform%
\footnote{We take the parameter of the Laplace transform as $ s = -i \omega $, in accordance with \onlinecite{leaver1986spectral, andersson1995excitation, berti2006quasinormal}. The Laplace transform $\hat{f}(\omega)$ of $f(t)$, defined in a certain interval $t_{0}< t< \infty$, is well defined if there exists $\lambda$, $\epsilon$ and $\tau$, such that
	\[
	f(t) < \lambda e^{\epsilon \tau}, \quad \forall t>\tau.
	\]
	Under these assumptions, $\hat{f}(\omega)$ exists for all $\omega \in \mathbb{C}$ such that $\operatorname{Im}(\omega)>\epsilon$. For a rigorous approach regarding the existence and unicity of the Laplace transform in Schwarzschild spacetime, we refer to \onlinecite{nollert1992quasinormal,andersson1995excitation,kay1987linear}.}  
\begin{align}
\Laplace\{\phi_{\ell}(t,r)\}=\hat{\phi}_{\ell}(\omega,r) = \int_{t_{0}}^{\infty} \phi_{\ell}(t,r)e^{i\omega t} dt.
\end{align}
Applying the Laplace transform to \eqref{eq:radial temporal equation}, we obtain an ordinary differential equation,
\begin{align}
P_{2}(\omega,r) \frac{\partial^{2}\hat{\phi}(\omega,r)}{\partial r^{2}} + P_{1}(\omega,r) \frac{\partial \hat{\phi}(\omega,r)}{\partial r} \nonumber\\
+ P_{0}(\omega,r) \hat{\phi}(\omega,r) = \mathcal{I}(\omega,r),
\label{eq:radial equation Laplace}
\end{align}
for each $\omega$, with $\mathcal{I}(\omega,r)$ taking care of the initial conditions. We omitted the index $ \ell $ to not clutter notation.

Eq.~(\ref{eq:radial equation Laplace}) can be rewritten as a Schr\"{o}dinger-type equation, 
\begin{align}
\frac{d^2 \hat{\psi}}{dr_{*}^2} - s(r_{*})\hat{\psi} = f(r_{*}),
\label{eq:radial equation Liouville}
\end{align}
by using a suitable change of variables,
\begin{align}
\hat{\phi} \to \hat{\psi}, \quad r \to r_{*}, 
\label{eq:Liouville coordinate transformation}
\end{align} 
which maps $r$ into an interval $(r^{min}_{*},r^{max}_{*})$. This is to be determined by the specific form of the metric. The solution of Eq.~\eqref{eq:radial equation Liouville} can then be found by the standard Green's function method and can be expressed as
\begin{align}
\hat{\psi}(\omega,r_{*}) = \frac{\hat{\psi}_{b}(\omega,r_{*})}{W[\hat{\psi}_{b},\hat{\psi}_{\infty}]} \int_{r_{*}}^{r^{max}_{*}} f(\omega,r_{*}')\hat{\psi}_{\infty}(\omega,r_{*}')dr_{*}'& \nonumber \\
+ \frac{\hat{\psi}_{\infty}(\omega,r_{*}) }{W[\hat{\psi}_{b},\hat{\psi}_{\infty}]}\int_{r^{min}_{*}}^{r_{*}} f(\omega,r_{*}')\hat{\psi}_{b}(\omega,r_{*}')dr_{*}'.
\label{eq:general Laplace solution}
\end{align}
Here, $W[\hat{\psi}_{b},\hat{\psi}_{\infty}]$ is the Wronskian of the solutions $\hat{\psi}_{b}$ and $\hat{\psi}_{\infty}$  of the homogeneous equation associated with \eqref{eq:radial equation Liouville},
\begin{align}
W_{r_{*}}[\hat{\psi}_{b},\hat{\psi}_{\infty}]=\hat{\psi}_{b}\frac{\partial \hat{\psi}_{\infty}}{\partial r_{*}} - \frac{\partial \hat{\psi}_{b}}{\partial r_{*}} \hat{\psi}_{\infty}.
\end{align}
The function $ \hat{\psi}_{b} $ should be determined after imposing some condition at $r^{min}_{*}$, deep into the bulk. This could be a regularity condition at the ``origin'' $r=0$ when $M= AdS$ or a condition at the event horizon when $M$ contains a black hole. On the other hand, the function $ \hat{\psi}_{\infty}$ is determined from the boundary conditions at the conformal infinity, $r^{max}_{*}$.

Assuming initial data with compact support, we find the following asymptotic approximation:
\begin{align}
\hat{\psi}(\omega,r_{*}) \approx \mathscr{A}(\omega) \hat{\psi}_{\infty}(\omega,r_{*}),  \mbox{  as } r \to r^{max}_{*},
\end{align}  
with $\mathscr{A}(\omega) =(1/W[\hat{\psi}_{b},\hat{\psi}_{\infty}]) \int_{r^{min}_{*}}^{r^{max}_{*}} f(\omega,r_{*}')\hat{\psi}_{b}(\omega,r_{*}')dr_{*}'$. 
Inverting the transformation \eqref{eq:Liouville coordinate transformation} leads to
\begin{align}
\hat{\phi}(\omega,r) \approx \mathscr{A}(\omega) \hat{\phi}_{\infty}(\omega,r), \mbox{ as } r \to \infty,
\end{align}
where $\hat{\phi}_{\infty}(\omega,r)$ is a solution of the homogeneous equation associated with \eqref{eq:radial equation Laplace} obeying some boundary condition at spatial infinity. The inverse Laplace transform then yields
\begin{align}
\phi(t, r)\approx \frac{1}{2\pi} \int_{-\infty + i \epsilon}^{+\infty + i\epsilon} \mathscr{A}(\omega) \hat{\phi}_{\infty}(\omega,r) e^{-i \omega t}d\omega,
\label{eq:Laplace approximation}
\end{align}
as $ r \to \infty$.

We note that the boundary conditions affect the resulting scalar field by means of the solutions of the homogeneous equation, $ \hat{\psi}_{b} (\omega, \rtort) $ and $ \hat{\psi}_{\infty} (\omega, \rtort) $, while the initial data are encoded in $ f (\omega, \rtort) $. Equivalently, the transformation \eqref{eq:Liouville coordinate transformation} allows one to interpret the dependence of the solution on the boundary conditions in terms of (the fundamental set of solutions of the homogeneous equation associated with \eqref{eq:radial equation Laplace}) $\{\hat{\phi}_{b}, \hat{\phi}_{\infty}\} $, while its dependence on the initial conditions is given by $ \mathcal{I}(\omega, r) $.

Since our aim here is to study the flux of energy at the conformal boundary, we will not fix any specific conditions on the field in the bulk other than requiring the usual regularity conditions, such as initial data with compact support and finiteness of the integrals associated with the asymptotic approximations. As a matter of fact, the convergence of these integrals depends on the analytical structure of the Green's function, which, in turn, depends on the boundary conditions deep inside the spacetime bulk. Hence, the convergence of these integrals must be treated differently for each spacetime. Throughout this work, we will assume that it is always possible to find an approximation such as \eqref{eq:Laplace approximation} for the spacetime at hand.

\section{Energy flux in asymptotically anti-de Sitter spacetimes}
\label{sec:Energy flux in asymptotically anti-de Sitter spacetimes}
We are now ready to study under what conditions the system \emph{spacetime} + \emph{field} is isolated, in the sense of having no energy flux through the timelike spatial boundary at infinity. As discussed in Sec. \ref{sec:Asymptotic behaviour of the field}, the asymptotic behavior of the solutions of \eqref{eq:wave equation general} is encoded in $ \hat{\phi}_{\infty, \ell} (\omega, r) $  (we, henceforth, reinsert the $ \ell $ index for definiteness). For each value of $ \ell $, this function satisfies the homogeneous equation associated with \eqref{eq:radial equation Laplace} in the limit $ r \to \infty $, which is given by
\begin{align}
\frac{\partial^2 }{\partial \rho^2} \hat{\phi}_{\infty,\ell}(\omega,\rho) + (n-2)\sec \rho \csc \rho \frac{\partial}{\partial \rho}  \hat{\phi}_{\infty,\ell}(\omega,\rho) \nonumber\\
+ \left[ \omega^{2} - \frac{\ell (\ell + n-3)}{\sin^{2}\rho} - \frac{\meff}{\cos^{2}\rho}  \right] \hat{\phi}_{\infty,\ell}(\omega,\rho) = 0,
\label{eq:AdSn equation of motion Laplace}
\end{align}
where we have changed the radial coordinate to $\rho$, with $r = \tan \rho$. Multiplying the last equation by $(\tan \rho)^{n-2}$ and  performing the transformation 
\begin{align}
\hat{\phi}_{\infty,\ell}(\omega,\rho) = \frac{Z_{\ell}(\omega,\rho)}{(\tan\rho)^{\frac{n-2}{2}}},
\label{eq:Liouville transformation}
\end{align} 
we find
\begin{align}
\frac{\partial^2 Z_{\ell}(\omega,\rho)}{\partial \rho^{2}} + \left[ \omega^{2} - V(\rho) \right] Z_{\ell}(\omega,\rho) = 0,
\label{eq:Schrodinger form}
\end{align}
where the effective potencial $V$ is given by 
\begin{align}
V(\rho)& = \left[\ell (\ell+n-3)+\frac{1}{4} \left(n^2-6 n+8\right)\right] \csc
^2\rho  \nonumber \\
& +\left[\frac{1}{4} n(n-2) + \meff \right] \sec ^2\rho.
\label{eq:potencial efetivo}
\end{align}

\noindent We also define
\begin{align}
d  = n-1, & &\nu^{2} = \frac{(n-1)^{2}}{4} + \meff,
\end{align}	
and
\begin{align}
a & = \frac{1}{2} \left( \frac{d}{2} + \ell + \nu - \omega \right), \\
b & = \frac{1}{2} \left( \frac{d}{2} + \ell + \nu + \omega \right).
\end{align}

\subsection{A convenient fundamental set of solutions}
\label{sec:fundamental_solutions}

For the sake of definiteness, let us fix a convenient set $\{Z_{\ell}^{(D)},Z_{\ell}^{(N)}\}$ of linearly independent solutions of \eqref{eq:Schrodinger form}. Following \onlinecite{gradshteyn2014table}, we take these functions as follows.
\begin{enumerate}
	\item[(i)] For $\nu$ not being an integer,
	\begin{align}
	Z_{\ell}^{(D)}(\omega,\rho) & = (\cos \rho)^{\frac{1}{2}+\nu} (\sin\rho)^{l+\frac{d-1}{2}} \times \nonumber \\
	& _2F_1\left(a,b;1+\nu;\cos ^2\rho \right), \label{eq:Z Dirichlet}\\
	Z_{\ell}^{(N)}(\omega,\rho) & = (\cos \rho)^{\frac{1}{2}-\nu } (\sin \rho)^{l+\frac{d-1}{2}} \times \nonumber\\
	& _2F_1\left(a - \nu ,b - \nu;1-\nu ;\cos^2 \rho \right). \label{eq:Z Neumann}
	\end{align}
	\item[(ii)] For $\nu=0$,
	\begin{align}
	Z^{(D)}_{\ell}(\omega,\rho) & = (\cos \rho)^{\frac{1}{2}+\nu} (\sin\rho)^{l+\frac{d-1}{2}} \times \nonumber \\
	& \quad _{2}F_1\left(a,b;1 ;\cos^2 \rho \right), \\
	Z^{(N)}_{\ell}(\omega,\rho) & = (\cos \rho)^{\frac{1}{2}+\nu} (\sin\rho)^{l+\frac{d-1}{2}} \times \nonumber \\
	&  \left\{ _{2}F_1\left(a,b;1 ;\cos^2 \rho \right) \ln(\cos^2 \rho) \right. \nonumber \\
	& + \sum_{k=1}^{\infty} \frac{(a)_{k}(b)_{k}}{(k!)^2} (\cos\rho)^{2 k} \times \left[\right. \psi(a+k)- \psi (a)   \nonumber\\ 
	&\left. \left.+ \psi(b+k) - \psi(b) -2 \psi(k+1) + 2 \psi(1) \right] \right\}.
	\end{align}
	\item[(iii)] For $\nu$ being a positive integer,
	\begin{align}
	Z^{(D)}_{\ell}(\omega,\rho) & = (\cos \rho)^{\frac{1}{2}+\nu} (\sin\rho)^{l+\frac{d-1}{2}} \nonumber \\
	& _{2}F_1\left(a,b;1 +\nu ;\cos^2 \rho \right), \\
	Z^{(N)}_{\ell}(\omega,\rho) & = (\cos \rho)^{\frac{1}{2}+\nu} (\sin\rho)^{l+\frac{d-1}{2}} \times \nonumber \\
	&  \left\{ _{2}F_1\left(a,b;1+\nu ;\cos^2 \rho \right) \ln(\cos^2 \rho) \right. \nonumber \\
	& + \sum_{k=1}^{\infty} \frac{(a)_{k}(b)_{k}}{(1+\nu)_{k} k!} (\cos\rho)^{2 k} \times \left[ h(k) - h(0) \right]  \nonumber\\
	&   \left.	- \sum_{k=1}^{\nu} \frac{(k-1)!(-\nu)_{k}}{(1-a)_{k}(1-b)_{k}} (\cos \rho)^{-2k} \right\},
	\end{align}	
\end{enumerate}
where
\begin{align}
\psi(x) & = \frac{d}{dx} \ln \Gamma (x), \\
h(k) & = \psi(a+k)+\psi(b+k) - \psi(1+\nu+k) \nonumber \\
& - \psi(k+1).
\end{align}
We note that, depending on the field mass $\mu$ and coupling constant $\xi$, the value of $\nu^{2}$ can be greater, less,  or equal to zero. With no loss of generality, we will consider $\nu > 0$ in the first case and $\nu = i \eta$, $\eta>0$, in the last case.

The general solution of \eqref{eq:Schrodinger form} can be written in terms of the fundamental solutions above as
\begin{align}
Z_{\ell} = \mathcal{N}_{\ell}\left[ \cos \zeta \, Z^{(D)}_{\ell} +\sin\zeta \, Z^{(N)}_{\ell} \right], 
\label{eq:Z general}
\end{align}
where $ \mathcal{N}_{\ell} $ does not depend on $ \rho $ and $ \zeta \in [0, \pi]$ does not depend neither on $ \rho $ nor on $ \ell $.
We will refer to the condition $ \zeta = 0 $ as the {\em generalized Dirichlet boundary condition} and to the function $ Z^{(D)}_{\ell} $ as the {\em Dirichlet solution}. The {\em generalized Neumann boundary condition} will be defined by $ \zeta = \pi/ 2 $, and we will refer to $Z^{(N)} _{\ell} $ as the {\em Neumann solution}.  The other values of $ \zeta \in [0, \pi] $ parametrize the {\em generalized Robin boundary conditions}.

As shown in Ref. \onlinecite{ishibashi2004dynamics}, the motivation for this terminology comes from the case of a conformally coupled field,  for which we have
\begin{align}
\mu ^{2} = 0, & \qquad \quad \xi = \frac{(n-2)}{4(n-1)}, & \nu =\frac{1}{2}.
\end{align}
In this case, the effective potencial \eqref{eq:potencial efetivo} is non-singular at $\rho = \pi/2$, and the ratio 
\begin{align}
\frac{ \partial Z_{\ell}/\partial \rho}{Z_{\ell}} \bigg|_{\rho\,  = \frac{\pi}{2}}
\label{eq:razao condicao contorno}
\end{align}
is well defined. The general solution \eqref{eq:Z general} can be written as
\begin{align}
Z_{\ell}(\rho) =   G_{\nu}(\rho) \left\{ \sin \zeta + \cos\zeta (\cos \rho)^{2\nu}  + \cdots \right\}
\end{align}
with
\begin{align}
G_{\nu \ell}(\rho) = \mathcal{N}_{\ell}\,(\cos\rho)^{-\nu + \frac{1}{2}} (\sin \rho)^{\frac{n-2}{2}+\ell}
\end{align}
so that the ratio \eqref{eq:razao condicao contorno} becomes 
\begin{align}
\frac{ \partial Z_{\ell}/\partial \rho}{Z_{\ell}} \bigg|_{\rho \, =\frac{\pi}{2}}=-\cot\zeta.
\label{eq:razao conforme}
\end{align}
We note that $ \zeta = 0 $ and $\zeta =\pi$ correspond to $ Z_{\ell} \big |_ {\rho = \pi / 2} = 0 $, which is the usual Dirichlet boundary condition. On the other hand, $ \zeta = \pi / 2 $ corresponds to $ \partial Z_{\ell} / \partial \rho \big |_{\rho = \pi / 2} = 0 $, the usual Neumann boundary condition. Other choices of $ \zeta \in [0, \pi] $ correspond to Robin boundary conditions. In the general case, the effective potential \eqref{eq:potencial efetivo} diverges as $ \rho$ goes to $ \pi / 2 $, and the ratio $ (dZ _ {\ell} / d \rho) / Z_{\ell} $ is no longer well defined. Despite that, the behavior of $ G ^ {-1} _ {\nu \ell} Z _ {\ell} $ as $ \rho $ goes to $\pi/2$ is dictated by $ \sin \zeta $, while the behavior of $ \partial (G ^ {- 1}_{\nu \ell}Z_ {\ell}) / \partial \rho $ is governed by $ \cos \zeta $, so it seems natural to define the ``generalized Dirichlet boundary condition'' as $\zeta = 0 $ and the ``generalized Neumann boundary condition'' by $ \zeta = \pi / 2 $. The other values of $ \zeta \in [0, \pi] $ parameterize the ``generalized Robin boundary conditions''.

\subsection{The flux at infinity}
\label{sec:The flux at infinity}

According to Weyl's {\em limit point} and {\em limit circle} theory, the allowed boundary conditions at the endpoints of the interval where a Sturm-Liouville problem is defined depend on the integrability of the solutions in the vicinity of these points. \cite{dappiaggi2016hadamard,zettl2005sturm} In the present case, the solutions of \eqref{eq:Schrodinger form} provide an approximation for the field near the point $ \rho = \pi / 2 $. The integrability of these solutions depends on the parameter $\nu$. 

In what follows, we are going to use the improved energy-momentum tensor of the complex scalar field, \cite{avis1978quantum,breitenlohner1982positive}
\begin{align}
T_{\alpha \beta}=  &   \frac{1}{2}\left(\partial_{\alpha} \Phi \, \partial_{\beta} \Phi^{*}+\partial_{\beta} \Phi \, \partial_{\alpha} \Phi^{*} \right) \nonumber\\
&- \frac{1}{2}g_{\alpha \beta} \left[g^{\rho \sigma} \partial_{\rho}\Phi \partial_{\sigma} \Phi^{*} +\meff \Phi\Phi^{*}\right] \nonumber \\
&+\xi \left(\ricci_{\alpha\beta} - g_{\alpha\beta} \Box - \nabla_{\alpha}\nabla_{\beta} \right)\Phi\Phi^{*},
\label{eq:AdS energy-momentum non minimal complex}
\end{align}	
to calculate the energy flux. The Killing vector field $k = \partial / \partial t$ gives rise to the formally conserved energy $Q^{\alpha} = |g|^{1/2}\, T^{\alpha\beta}k_{\beta}$ $(\partial_{\mu} Q^{\mu} = 0)$, and the energy flux across the  spatial infinity is  given by%
\begin{align}
\flux = -\lim\limits_{\rho \to \pi/2 } \int d\theta_{1}\dots d\theta_{n-2}\, \, g^{\rho\rho} \, Q_{\rho}.
\label{eq:flux infinity}
\end{align}

\subsubsection*{The case $\nu^{2}\ge 1$}
This is a simplest instance to analyze. In this case, $ Z_ {\ell}^{(D)} $ is square integrable near $ \rho = \pi / 2 $, while $ Z_ {\ell}^{(N)} $ is not. As a result, the generalized Dirichlet boundary condition must be chosen in this case.  With this boundary condition the energy flux across the  spatial infinity turns out to be zero. We omit the calculation since it is identical to the case $0<\nu^2<1$, treated below, once we set $\zeta=0$.

\subsubsection*{The case $0<\nu^{2}<1$}
In this case, both solutions are square integrable near $ \rho = \pi / 2 $. The allowed boundary conditions are therefore of Robin type. 

For these values of $\nu$, \eqref{eq:Z general} and \eqref{eq:Liouville transformation} imply the following asymptotic behavior for $ \hat {\phi}_{\infty, \ell} $:
\begin{align}
\hat{\phi}_{\infty,\ell}(\omega,\rho) \approx \mathcal{N}_{\ell}(\omega) \left[ \cos \zeta \, \hat{\phi}^{(D)}_{\ell}(\omega,\rho)  \right.\nonumber \\
\left.
+ \sin \zeta \, \hat{\phi}^{(N)}_{\ell}(\omega,\rho) \right] ,
\label{eq:AdSn radial asymptotic Laplace}
\end{align}
where 
\begin{align}
\hat{\phi}^{(D)}_{\ell}\left(\omega,\rho\right) = \left( \frac{\pi}{2} - \rho \right)^{ \frac{d}{2}+\nu} + J_{\ell}(\omega)\left( \frac{\pi}{2} - \rho \right)^{ \frac{d}{2}+\nu +2} \nonumber\\ 
+ O\left[ \left( \frac{\pi}{2} - \rho \right)^{\frac{d}{2}+\nu +4} \right], \\
\hat{\phi}^{(N)}_{\ell}\left(\omega,\rho\right) = \left( \frac{\pi}{2} - \rho \right)^{ \frac{d}{2} - \nu} + K_{\ell}(\omega)\left( \frac{\pi}{2} - \rho \right)^{ \frac{d}{2}-\nu +2} \nonumber\\
+ O\left[ \left( \frac{\pi}{2} - \rho \right)^{\frac{d}{2}-\nu +4} \right], 
\label{eq:AdSn radial asymptotic Dirichlet Neumann}
\end{align}
with
\begin{align}
J_{\ell}(\omega) & = \frac{a_{1}(\omega) b_{1}(\omega)}{1+\nu} - \frac{n -1 + 6 \ell + 2 \nu}{12}, \\
K_{\ell}(\omega) & = \frac{a_{2}(\omega)b_{2}(\omega)}{1-\nu} - \frac{n-1 + 6 \ell -2\nu}{12},
\end{align}
as $\rho\to \pi/2 $.
Upon substitution of \eqref{eq:AdSn radial asymptotic Laplace} into \eqref{eq:Laplace approximation}, we obtain the following asymptotic expression for $\phi_{\ell}$:
\begin{align}
\phi_{\ell}(t,\rho )& \approx  \cos \zeta \left( \frac{\pi}{2} - \rho \right)^{\frac{d}{2}+\nu} \mathcal{T}_{\ell}(t)  \nonumber \\
& + \cos \zeta \left( \frac{\pi}{2} - \rho  \right)^{\frac{d}{2}+\nu+2} \mathcal{T}_{D,\ell}(t) \nonumber\\
& + \sin \zeta \left( \frac{\pi}{2} - \rho \right)^{\frac{d}{2}-\nu} \mathcal{T}_{\ell}(t)\nonumber\\
&  +\sin \zeta \left( \frac{\pi}{2} - \rho \right)^{\frac{d}{2}-\nu + 2} \mathcal{T}_{N,\ell}(t),
\label{eq:radial approximation}
\end{align}
where
\begin{align}
\mathcal{T}_{\ell}(t) & =\frac{1}{2 \pi} \int_{-\infty + i \epsilon}^{+\infty + i \epsilon} \mathscr{A}_{\ell}(\omega)\mathcal{N}_{\ell}(\omega) e^{-i \omega t} d\omega, \\
\mathcal{T}_{D,\ell}(t) & =\frac{1}{2 \pi} \int_{-\infty + i \epsilon}^{+\infty + i \epsilon} \mathscr{A}_{\ell}(\omega)\mathcal{N}_{\ell}(\omega)J_{\ell}(\omega) e^{-i \omega t}d\omega, \\
\mathcal{T}_{N,\ell}(t) & =\frac{1}{2 \pi} \int_{-\infty + i \epsilon}^{+\infty + i \epsilon} \mathscr{A}_{\ell}(\omega)\mathcal{N}_{\ell}(\omega)K_{\ell}(\omega) e^{-i \omega t} d\omega.	
\end{align}

Using the asymptotic form \eqref{eq:radial approximation}, \eqref{eq:AdS energy-momentum non minimal complex} and \eqref{eq:flux infinity}, we get
\begin{align}
\flux \sim \lim\limits_{\rho \to \pi/2} \sin \zeta \left\{\cos\zeta \,  A + \sin \zeta \, B \left( \frac{\pi}{2} - \rho \right)^{-2\nu}  \right\} \times \nonumber \\
\left( \sum_{\{\ell\}} \frac{d}{dt}|\mathcal{T}_{\ell}(t)|^{2} \right),
\label{eq:flux 0<nu^2<1}
\end{align}
where
\begin{align}
&A =  \frac{d}{2}- 2 \xi (d+1), \label{eq:A 0<nu^2<1}\\
&B =  \left(\frac{1}{4}-\xi \right)(d-2 \nu) -\xi  \label{eq:B 0<nu^2<1}.
\end{align}

We immediately see that by imposing the Dirichlet boundary condition ($ \zeta = 0 $) the flow of energy across the   infinity turns out to be zero. 

On the other hand, when $ \zeta \neq 0 $, we must choose the coupling constant so that $ B = 0 $ in order that the energy flux be finite. This leads to
\begin{align}
\flux \sim \sin \zeta \cos \zeta \, A	 \sum_{\{\ell\}} \frac{d}{dt}|\mathcal{T}_{\ell}(t)|^{2}.
\label{eq:energy flux}
\end{align}	
The integrals defining $ \mathcal {T}_ {\ell} (t)$, $ \ell = 0, 1, 2 \dots $, depend on the singularity structure of the functions $ \mathscr{A}_{\ell}$, which, in turn, depend on the boundary conditions in the spacetime bulk and on the initial conditions of the system. As a result, except for very specific field configurations, we must impose the Neumann condition ($ \zeta = \pi / 2 $) for the system to become effectively isolated. For general Robin conditions ($ \zeta \ne 0 $ and $ \zeta \ne \pi / 2 $), the energy flux across the conformal boundary is generically not zero.%

There is a notable case where the energy flux \eqref{eq:energy flux} can be zero without imposing either $\zeta = 0$ or $\zeta=\pi/2$. For the propagation of a single mode of frequency $\omega \in \mathbb{R}$, we have $\mathcal{T}_{\ell}(t) \sim e^{-i \omega t}$, and then, $d|\mathcal{T}_{\ell}(t)|^{2}/dt=0$, and the energy flux \eqref{eq:energy flux} vanishes for every Robin boundary condition $\zeta \in [0, \pi]$. However, for the propagation of two field modes, this conclusion is no longer true. More generally, if the scalar field is composed of a nontrivial superposition of modes of different frequencies, then $d|\mathcal{T}_{\ell}(t)|^{2}/dt\neq0$.

In summary, the boundary conditions that make the system {\em scalar field} + {\em spacetime} effectively isolated in this case are as follows:
\begin{enumerate}
	\item[(i)] $\zeta = 0$ (Dirichlet).
	\item[(ii)] $\zeta=\pi/2$ (Neumann), together with $\xi$ chosen  such that $B=0$.
\end{enumerate}
In particular, for a minimally coupled field ($\xi =0$), only the Dirichlet boundary condition gives zero energy flux across the  spatial infinity since in this case, $B \ne 0$.

\subsubsection*{The case $\nu^{2}=0$}
As in the previous case, both solutions are square integrable near $ \rho = \pi / 2 $ here. The allowed boundary conditions are therefore again of Robin type. 

Moreover, the behavior of both $G_{\nu\ell}^{-1}Z_\ell$ and $\partial(G_{\nu\ell}^{-1}Z_\ell))/\partial\rho$ are governed by $\sin \zeta$ for $\nu=0$. Thus, one can interpret $\zeta = 0$ (or $\zeta = \pi$) as the simultaneous imposition of generalized Neumann and Dirichlet boundary conditions. Following the same steps as in the previous case, we find that the condition of zero flux again requires $\zeta = 0$ together with $\xi = (n-1)/4n$. 

\subsubsection*{The case $\nu^2<0$}
We now consider the case when $\nu^{2}<0$, i.e., $\nu = i \eta$ with $\eta>0$. Once again, both solutions are square integrable near $\rho=\pi/2$ here. The energy flow across the  spatial infinity is now given by
\begin{align}
\mathcal{F} \approx \lim\limits_{\rho \to \pi/2} \sum_{\{\ell\}}\left(A_{\ell} \cos2\zeta + B_{\ell} \sin 2\zeta + C_{\ell}\right) ,
\label{eq:flux nuQuad negativo}
\end{align}
where
\begin{align}
A_{\ell} & = \eta \operatorname{Im}\left\{ \mathcal{T}^{*}_{\ell}(t) \frac{d\mathcal{T}_{\ell}(t)}{dt} \right\} , \label{eq:A nu^2<0}\\
B_{\ell} & = \frac{1}{2}  \operatorname{Re}\left\{ \left[ (n+2i\eta)(4\xi - 1) + 1 \right] \left(\frac{\pi}{2} - \rho \right)^{2 i \eta} \right\}\times \nonumber\\
& \qquad \operatorname{Re} \left\{ \mathcal{T}^{*}_{\ell}(t) \frac{d\mathcal{T}_{\ell}(t)}{dt} \right\}, \label{eq:B nu^2<0}\\
C_{\ell} & = \frac{1}{2} \left[ 1+n(4\xi -1) \right] \operatorname{Re} \left\{ \mathcal{T}^{*}_{\ell}(t) \frac{d\mathcal{T}_{\ell}(t)}{dt} \right\}. \label{eq:C nu^2<0}
\end{align}
Since the functions $\sin 2\zeta$ and $\cos 2\zeta$	are linearly independent, we conclude that, in general, the system cannot be treated as isolated for $\nu^{2}<0$. 

Once again, a notable exception is given by the propagation of a single mode with frequency $\omega \in \mathbb{R}$. In this case, we have $\mathcal{T}(t) \sim e^{i \omega t}$, and therefore, $\operatorname{Re} \left\{ \mathcal{T}^{*}(t) [d\mathcal{T}(t)/dt] \right\}=0$, which implies that the coefficients $B$ and $C$ in \eqref{eq:flux nuQuad negativo} both vanish. Then, by choosing the boundary condition as $\zeta = \pi /4$, we can cancel out the energy flux through the conformal boundary.

It is worth mentioning that when $M$ is not only \emph{asymptotically} AdS, but $M=AdS$, the differential operator associated with equation \eqref{eq:Schrodinger form} is unbounded below for $ \nu^{2} < 0 $. As a result, one cannot find positive self-adjoint extensions of it \cite{ishibashi2004dynamics} so that it is not possible to define a physically ``reasonable'' time evolution in this case.%
\footnote{By ``reasonable'', we mean precisely the conditions given in Ref. \onlinecite{wald1980dynamics,ishibashi2003dynamics}.} In general, one cannot make assertions concerning the positivity of the differential operator associated with the correspondent radial equation without detailed information about the bulk structure of spacetime. Indeed, the positivity of the differential operator may be somewhat subtle to be rigorously established even when the bulk structure is fully known. \cite{garbarz2017scalar}

Finally, we note that the calculations in this section could be performed using the canonical (non-improved) energy-momentum tensor,
\begin{align}
\widetilde{T}_{\alpha \beta}=  &   \frac{1}{2}\left(\partial_{\alpha} \Phi \, \partial_{\beta} \Phi^{*}+\partial_{\beta} \Phi \, \partial_{\alpha} \Phi^{*} \right) \nonumber\\
&- \frac{1}{2}g_{\alpha \beta} \left[g^{\rho \sigma} \partial_{\rho}\Phi \partial_{\sigma} \Phi^{*} +\meff \Phi\Phi^{*}\right].
\label{eq:AdS energy-momentum non improved complex}
\end{align}
In this case, we find that (i) for $\nu^{2}>0$, only the Dirichlet boundary condition yields a zero energy flux across infinity and (ii) for $\nu^2 \le 0$, the flux is generically nonzero even for the Dirichlet choice. These results are also what one would obtain by formally substituting $\xi=0$ in the above calculations for the improved energy-momentum tensor.

\subsection{Mode analysis}
To conclude this section, we discuss how our results fit with the existing literature. A common approach consists in considering a time dependence given by $ e ^ {- i \omega t} $ and to impose boundary conditions on the radial part for each field mode of frequency $\omega$. \cite{dappiaggi2016hadamard,dappiaggi2018superradiance,ferreira2017stationary,morley2020quantum,barroso2020boundary} 
For simplicity and in order to make the discussion clearer, let us consider the specific case of $n=3$, i.e., of an asymptotically AdS$_{3}$ spacetime  . 

The allowed values of $ \omega $ for the field eigenfunctions are determined from the boundary conditions in the bulk and at infinity, with $ \omega \in \mathbb {R} $ or $ \omega \in \mathbb{C} $, depending on the specific conditions imposed. In the following, we will consider both the energy flow due to the propagation of a single frequency mode $\omega_{1}$ and the flux due to the propagation of a superposition of modes with frequencies $ \omega_{1}$ and $\omega_ {2} $. Since we are not imposing any boundary condition on the spacetime bulk, we will allow $\omega$ to be complex and then specialize to the case of a real $\omega$.

Let us consider the case when $ 0< \nu <1$. Let $\Phi_{1}$ be a mode with frequency $ \omega_{1} \in \mathbb{C} $,\
\begin{align}
\Phi_{1}(t,\rho,\varphi) = \phi_{\omega_{1}\ell}(\rho) \,  e^{- i \omega_{1} t} e^{i \ell \varphi}.
\end{align}
A straightforward calculation shows that the energy flux across infinity for this specific solution is given by
\begin{align}
\mathcal{F}^{(1)}& \sim \lim_{\rho \to \pi/2} e^{2 \operatorname{Im}(\omega_{1}) \,  t } \operatorname{Im}(\omega_{1}) \sin \zeta \, \,  \times \nonumber\\
& \left\{ \cos \zeta \, (1-6\xi) + \sin\zeta \, B \left( \frac{\pi}{2} - \rho \right)^{-2 \nu} \right\},
\label{eq:fluxo complexo 1 modo}
\end{align} 
where 
\begin{align}
B = \left[ \left( \frac{1}{2} - 2 \xi  \right)(1-\nu) -\xi \right].
\end{align}
We immediately see that when $ \operatorname{Im} (\omega_{1}) \ne 0 $, only the Dirichlet and Neumann boundary conditions cancel the flux (the latter with $ \xi $ chosen such that $ B = 0 $ as usual). On the other hand, when $\omega \in \mathbb{R}$, the energy flux is null for any Robin boundary condition ($ 0 \le \zeta <\pi $), regardless of the coupling constant $ \xi $. However, this is only a result of the very particular situation of a single mode solution. Considering the superposition of even just two modes, $ \Phi_{1} $ and $ \Phi_{2} $, with frequencies $ \omega_{1}, \, \omega_ {2} \in \mathbb{R} $, we obtain the corresponding flux given by
\begin{align}
\mathcal{F}^{(1,2)} \sim \lim_{\rho \to \pi/2} \sin(\Delta \omega \, t) \, \sin\zeta   
\left\{ \cos\zeta \, (1-6\xi) \right. \nonumber \\
\left. + \sin\zeta  \, B \left( \frac{\pi}{2} - \rho \right)^{-2 \nu} \right\},
\label{eq:superposition flux}
\end{align} 
where $ \Delta \omega = \omega_{1} - \omega_{2} $. Therefore, once again, only the Dirichlet and Neumann boundary conditions are compatible with the hypothesis that the system is isolated  (the latter with $ \xi $ chosen such that $ B = 0 $ as usual). 

The analysis of the other cases of $\nu$ leads to the same conclusion. A single mode of real frequency can have zero flux at infinity while obeying Robin boundary conditions. However, as soon as we consider a superposition of modes of different real frequencies (or even a single mode of complex frequency), generic Robin boundary conditions are not compatible with zero flux at infinity and the results of Subsection \ref{sec:The flux at infinity} are recovered.

For the sake of completeness, we repeat this analysis for the case of a real scalar field in Appendix \ref{apendice:real}. The results are essentially the same, the only difference being that general Robin boundary conditions are not compatible with zero energy flux at infinity even in the case of a single mode.

We conclude this section by noting that our results do not depend on the bulk structure of the spacetime. Regardless of the bulk, the only boundary conditions at infinity that make the system effectively isolated are those of Dirichlet and Neumann types. Some particular field configurations may, of course, have zero flux without conforming to this rule. 
This is the case of a single mode of a complex scalar field with real frequency, for which the flux is zero irrespective of the choice of $\zeta$.

\section{Discussion}\label{sec:Discussion}	
We have studied the asymptotic behavior of scalar fields in spacetimes which are asymptotically anti-de Sitter. We determined the boundary conditions at the spatial infinity for which there is no flow of energy at the conformal boundary.  
We showed that the only allowed choices that are consistent with this requirement are the generalized Dirichlet and Neumann boundary conditions (the latter with a specific choice of the coupling constant). This happens regardless of the theory in the spacetime bulk. The energy flux was calculated using the improved energy-momentum tensor \eqref{eq:AdS energy-momentum non minimal complex}. If we had used the canonical energy-momentum tensor \eqref{eq:AdS energy-momentum non improved complex} instead, only the Dirichlet boundary conditions would be compatible with zero flux at the conformal boundary. 

In particular, Robin mixed boundary conditions, as considered, for instance, in Refs. \onlinecite{ishibashi2004dynamics} and \onlinecite{dappiaggi2016hadamard,dappiaggi2018superradiance,ferreira2017stationary,morley2020quantum,barroso2020boundary} (although physically reasonable since they provide a fully deterministic dynamics), are not compatible with the requirement that the spacetime is an isolated system. 

The case of an asymptotically AdS$_{2}$ spacetime  can be treated in a similar manner. The fundamental difference is that in this case, the spatial infinity has two distinct components so that, in order for the system to be isolated, one must demand the energy flow to be (separately) zero at each of the boundaries. We must then impose two independent conditions at each of the two boundaries. The zero flux condition constrains those to be, again, of Dirichlet and Neumann types.

\begin{acknowledgments}
R.A.M. acknowledges discussions with L. de Souza Campos. C.C.d.O. acknowledges support from the Conselho Nacional de Desenvolvimento Cient\'{i}fico e Tecnol\'{o}gico (CNPq, Brazil), Grant No. 142529/2018-4. R.A.M. was partially supported by Conselho Nacional de Desenvolvimento Científico e Tecnológico under Grant No. 310403/2019-7. J.P.M.P. was partially supported by Conselho Nacional de Desenvolvimento Científico e Tecnológico under Grant No. 311443/2021-4.

\end{acknowledgments}

\appendix

\section{Principal and non-principal solutions}
\label{apendice:solprincipal}

For $\nu\in\mathbb{R}$, the function $Z^{(D)} _ {\ell} $ defined in Sec.~\ref{sec:fundamental_solutions} is the only solution (up to a multiplicative factor) such that
$
\lim_{\rho \to \pi/2}\left[ Z^{(D)}_{\ell}(\rho)/Z_{\ell}(\rho)\right] = 0
$ 
for any solution $ Z_{\ell} $ not proportional to $ Z^{(D)}_{\ell}$. A solution satisfying this condition is called a {\em principal} solution (at the endpoint $ \rho = \pi / 2$). Solutions that are not proportional to $ Z^{(D)}_ {\ell} $ are called {\em non-principal} (at the endpoint $ \rho = \pi / 2$). 

We note that non-principal solutions are not unique. In fact, if $ \tilde {Z} _ {\ell} $ is a non-principal solution, then $ \tilde{Z}_{\ell} + \alpha Z^{(D)}_{\ell} $ is also a solution of this type for any $ \alpha \in \mathbb{R}$. It is interesting to ask what would change in our analysis if we replace $ Z^{(N)}_ {\ell} $ of Sec.~\ref{sec:fundamental_solutions} by another non-principal solution $ \tilde{Z}^{(N)} = Z^{(D)} + \gamma Z^{(N)} $, $\gamma \in \mathbb {R} $.
In terms of the new set $ \{Z^{(D)}, \tilde{Z}^{(N)}\} $, the general solution of \eqref{eq:Schrodinger form} can be expressed as
\begin{align}
Z_{\ell} = \mathcal{N}_{\ell} \left[\cos\zeta \, Z_{\ell}^{(D)} + \sin \zeta \, \tilde{Z}^{(N)}_{\ell}\right],
\end{align}
and the condition $ \zeta = \pi / 2 $ no longer selects the function given in \eqref{eq:Z Neumann}. The value of $ \zeta $ that selects that function is now 	
\begin{align}
\cot \bar{\zeta} = -\gamma. 
\end{align}

The energy flux calculated in terms of the new set of solutions is given by 
\begin{align}
\flux \sim \lim\limits_{\rho \to \pi/2} \sin \zeta \left\{(\cos\zeta + \gamma \sin\zeta) \,  A +  \right. \nonumber \\
\left. \sin \zeta \, B \left( \frac{\pi}{2} - \rho \right)^{-2\nu}  \right\} \times \nonumber \\
\left( \sum_{\{\ell\}} \frac{d}{dt}|\mathcal{T}_{\ell}(t)|^{2} \right).
\label{eq:fluxo Neumann modificada}
\end{align}
From \eqref{eq:fluxo Neumann modificada}, we see that the boundary conditions that cancel the energy flux across the conformal boundary are $ \zeta = 0 $ (Dirichlet) and $ \zeta = \bar {\zeta} $ (along with $ \xi $ chosen such that $ B = 0 $). Therefore, regardless of how the generalized Neumann condition is defined, the boundary conditions associated with zero flux at infinity are those that select the solutions $ Z^{(D)} $ and $ Z^{(N)} $ of Sec.~\ref{sec:fundamental_solutions}.

\section{Real scalar fields}
\label{apendice:real}

We discuss in this appendix the behavior of the energy flux across the spatial infinity for real scalar fields. The improved energy-momentum tensor in this case is given by	 
\begin{align}
T_{\alpha \beta}  & =  \partial_{\alpha} \Phi \, \partial_{\beta} \Phi - \frac{1}{2}g_{\alpha \beta} \left[g^{\rho \sigma} \partial_{\rho}\Phi \partial_{\sigma} \Phi +\meff \Phi^{2} \right] \nonumber\\
&+\xi \left(\ricci_{\alpha\beta} - g_{\alpha\beta} \Box - \nabla_{\alpha}\nabla_{\beta} \right)\Phi^{2}.
\label{eq:AdS energy-momentum non minimal real}
\end{align}	 
The counterparts for real scalar fields of the real and complex frequency cases of the main text are, respectively, given as follows
\begin{enumerate}
	\item[(i)] $\cos (\omega t + \delta)$ when $\omega \in \mathbb{R}$;
	\item[(ii)] $e^{\omega_{I} t } \cos (\omega_{R} t + \delta)$ when $ \omega = \omega_{R} + i \omega_{I} \in \mathbb{C}$.
\end{enumerate}

Let us consider case (i) separately. Let $\Phi_{1}$ be a mode with frequency $\omega_{1} \in \mathbb{R}$,
\begin{align}
\Phi_{1}(t,\rho,\varphi)  & = \phi_{\omega_{1} \ell}(\rho) \cos(\omega_{1} t +\delta_{1} ) \times \nonumber \\
& \quad[C_{1} \cos \ell\varphi + D_{1} \sin\ell \varphi ] .
\end{align}
This leads to
\begin{align}
\phi_{\omega_{j} \ell}(\rho) \approx \cos \zeta \phi^{(D)}_{\omega_j \ell}(\rho) + \sin \zeta \phi^{(N)}_{\omega_j \ell}(\rho),
\end{align}
$j =1, 2,$ as $\rho \to \pi/2$, where
\begin{align}
\phi^{(D)}_{\omega_{j}\ell} (\rho)= (\sin \rho )^{\ell} (\cos \rho )^{1 + \nu} \, _2F_1\left(a_1,b_1;c_1;\cos ^2\rho \right), \\
\phi^{(N)}_{\omega_{j}\ell} (\rho)= (\sin \rho )^{\ell} (\cos \rho )^{1-\nu } \, _2F_1\left(a_2,b_2;c_2 ;\cos^2 \rho \right),
\end{align}
and
\begin{align}
a_1  & = \frac{1}{2} \left( 1 + \ell + \nu - \omega_{1} \right), \\
b_1  & = \frac{1}{2} \left( 1 + \ell + \nu + \omega_{1} \right),\\
c_1  & = 1+\nu, \\
a_2  & = \frac{1}{2} \left( 1 + \ell - \nu - \omega_{2} \right),\\
b_2  & = \frac{1}{2} \left( 1 + \ell - \nu + \omega_{2} \right), \\
c_2  & = 1-\nu.
\end{align}

The energy flux across the  spatial infinity is then given by
\begin{align}
\mathcal{F} & \approx	\omega_{1} \sin[2(\omega_{1} t + \delta_{1})] \sin\zeta \lim_{\rho \to \pi/2}\left\{ \cos\zeta \, (1-6\xi)  \right. \nonumber\\
&			\qquad \left.	+ \sin\zeta \, B \left( \frac{\pi}{2} - \rho \right)^{-2 \nu}	\right\},
\label{eq:fluxo campo real freq 1 modo}
\end{align}
and we see that this is zero only for the Dirichlet boundary condition ($ \zeta = 0 $) or the Neumann boundary condition ($ \zeta = \pi / 2 $) with $ \xi $ such that $ B = 0 $. This should be compared to the corresponding result for the complex field, Eq.~\eqref{eq:fluxo complexo 1 modo}, for which the flux associated with a single mode was found to be zero even for Robin conditions.

Now, consider the superposition of two modes (still in case (i)), $ \Phi_{1}$ and $ \Phi_{2} $, with
\begin{align}
\Phi_{1}(t,\rho,\varphi)  & = \phi_{\omega_{1} \ell}(\rho) \cos(\omega_{1} t +\delta_{1} ) \times \nonumber \\
& \quad[C_{1} \cos \ell\varphi + D_{1} \sin\ell \varphi ],  \\
\Phi_{2}(t,\rho,\varphi)  & = \phi_{\omega_{2} \ell}(\rho) \cos(\omega_{2} t +\delta_{2} ) \times \nonumber \\
& \quad [C_{2} \cos \ell\varphi + D_{2} \sin\ell \varphi ],
\end{align}
with $\omega_{1}, \omega_{2} \in \mathbb{R}$.
The energy flow across the conformal infinity is now given by
\begin{align}
\mathcal{F} \sim & [ \cos(\omega_{1} t + \delta_{1}) + \cos(\omega_{2} t + \delta_{2}) ] \times \nonumber\\
& \left[ \omega_{1} \sin(\omega_{1} t + \delta_{1}) + \omega_{2} \sin(\omega_{2} t + \delta_{2}) \right] 2 \sin\zeta \times \nonumber \\
&\lim_{\rho \to \pi/2} \left\{ \cos\zeta \, (1-6\xi) + \sin\zeta \, B \left( \frac{\pi}{2} - \rho \right)^{-2\nu} \right\}.
\label{eq:fluxo campo real freq real 2 modos}
\end{align} 
Since the functions $\sin \omega_{j} t$ and $\cos\omega_{j} t$ are linearly independent, the only boundary conditions that do not violate the isolated system hypothesis are again of the Dirichlet and the Neumann types (the latter with $ \xi $ such that $ B = 0 $).

We end by considering case (ii). The real scalar field mode in this case (the counterpart of the complex mode with complex frequency) is given by
\begin{align}
\Phi_{1}(t,\rho,\varphi)  & = \operatorname{Re}\left[  \phi_{\omega_{1} \ell}(\rho) \right] e^{\beta_{1} t} \cos(\alpha_{1} t +\delta_{1} ) \times \nonumber \\
& \quad[C_{1} \cos \ell\varphi + D_{1} \sin\ell \varphi ].
\end{align}
The energy flux through infinity is now
\begin{align}
\mathcal{F} & \approx e^{2 t \beta_{1} } \left[ \beta_{1} \cos^{2}(\alpha_{1} t + \delta_{1}) \, - \right. \nonumber\\ 
& \qquad \quad \left. \alpha_{1} \cos(\alpha_{1} t + \delta_{1}) \sin(\alpha_{1} t + \delta_{1})  \right] 2 \sin \zeta \times \nonumber \\
& \qquad \qquad \lim_{\rho \to \pi/2} \left\{ \cos \zeta (1-6\xi) + \sin\zeta \, B \left( \frac{\pi}{2} - \rho \right)^{-2\nu} \right\},
\end{align}
and we see again that the only conditions compatible with the hypothesis that the system is isolated are those of Neumann and Dirichlet (the latter with $ \xi $ such that $ B = 0 $).

\end{document}